\documentclass[%
 reprint,
 amsmath,amssymb,
 aps,
]{revtex4-2}

\usepackage{graphicx}
\usepackage{dcolumn}
\usepackage{bm}
\usepackage{ulem}
\usepackage{xcolor}

\begin{document}

\preprint{APS/123-QED}

\title{Plug-and-play measurement of chromatic dispersion by means of two-photon interferometry}

\author{Romain Dalidet$^1$, Anthony Martin$^1$, Mattis Riesner$^1$, Sidi-Ely Ahmedou$^2$, Romain Dauliat$^2$, Baptiste Leconte$^2$, Guillaume Walter$^3$, Grégory Sauder$^1$, Jean-Christophe Delagnes$^3$, Guy Millot$^{4,5}$, Philippe Roy$^2$, Raphaël Jamier$^3$, Sébastien Tanzilli$^1$, Laurent Labonté$^1$}
\email{laurent.labonte@univ-cotedazur.fr}
\affiliation{$^1$Université Côte d’Azur, CNRS, Institut de physique de Nice, France\\
$^2$Université de Limoges, CNRS, XLIM, UMR 7252, F-87000 Limoges, France\\
$^3$CELIA, Centre Lasers Intenses et Applications, Université de Bordeaux-CNRS-CEA, UMR 5107, F-33405 Talence Cedex, France\\
$^4$ICB, Université de Bourgogne, CNRS, UMR 6303, F-21078 Dijon, France\\
$^5$Institut Universitaire de France (IUF), 1 Rue Descartes, Paris, France
}

\begin{abstract}
 Since the first proof-of-principle experiments 25 years ago, quantum metrology has matured from fundamental concepts to versatile and powerful tools in a large variety of research branches, such as gravitational-wave detection, atomic clocks, plasmonic sensing, and magnetometry. At the same time, two-photon interferometry, which underpins the possibility of entanglement to probe optical materials with unprecedented levels of precision and accuracy, holds the promise to stand at the heart of innovative functional quantum sensing systems. We report a novel quantum-based method for measuring the frequency dependence of the velocity in a transparent medium, \textit{i.e}, the chromatic dispersion (CD). This technique, using energy-time entangled photons, allows  straightforward access to CD value from the visibility of two-photon fringes recorded in a free evolution regime. In addition, our quantum approach features all advantages of classical measurement techniques, \textit{i.e}, flexibility and accuracy, all in a plug-and-play system. 
 \end{abstract}

\maketitle

\section{Introduction}
Quantum metrology is one of the most advanced applications of quantum technologies exploiting quantum physics foundations towards real-world applications. While the fragility of quantum states poses stringent constraints to the development of quantum computer and quantum communication systems~\cite{Wei22, Fla19}, it allows unexpected possibilities for measurement systems.
Several physical platforms can be employed to develop quantum sensors as cold-atoms, nitrogen-vacancy centers or superconducting circuits, making precise measurements of time, accelerations, rotations, gravity, and magnetic fields~\cite{NV, geiger, taylor_high-sensitivity_2008, degen_quantum_2017, clarke_superconducting_2008}, respectively. They share as common feature on quantum properties, standing as a distinct and rapidly growing branch of research within the area of quantum science and technology. Another appealing quantum system are photons, enabled by their inherent properties such as high mobility, together with the available technology for their generation, manipulation, and detection. Recent development of platforms and techniques to generate suitable quantum photonic states  able to provide quantum-enhancement in different metrology tasks, such as biological systems~\cite{taylor_quantum_2016}, optical coherent tomography~\cite{abouraddy_quantum-optical_2002}, microscopy and imaging~\cite{kolobov_spatial_1999}, is of significant importance. Photonic quantum correlations represent a remarkable resource allowing to enhance the performance of quantum sensors~\cite{clark_special_2021}.\\

Since many physical problems can be regarded as phase estimation processes, two-photon interferometry represents one of the most relevant approach, notably for the measurement of dispersive properties such as the chromatic dispersion (CD)~\cite{rarity_high_visibility_1993}. The measurement of CD holds significant relevance in numerous applications within the field of photonics, particularly in classical~\cite{agrawal_nonlinear_1995} and quantum communications~\cite{fasel_quantum_2004}, as well as the generation of nonlinear effects~\cite{agrawal_nonlinear_1995}, encompassing platforms based on $\chi^{(2)}$ and $\chi^{(3)}$ nonlinearities. To date, measuring CD has been achieved using two main categories of techniques, relying on spectral or temporal properties of the light probe. On one hand, temporal CD measurements exhibit a great flexibility but requires using a spectrally broad light source and a picosecond time resolution detector over a kilometer-length, resulting in a moderate accuracy~\cite{Thevenaz}. On the other hand, interferometric techniques enable measurements with an excellent accuracy over widest spectral range, at the price of (i) complex experimental setup including spectrometer (ii) systematic errors, and (iii) moderate signal-to-noise ratio~\cite{Disp}. Even if quantum-enhancement of accuracy has recently been demonstrated, still, process is rather tedious, time-consuming and difficult to automate~\cite{kaiser_quantum_2018}.\\

In this work, we propose a disruptive approach among current CD measurement methods merging the best aspects of classical techniques with unique quantum features, in a plug-and-play fashion system associated with a 1\% accuracy. This method relies on the peculiar properties of photonic entanglement through two-photon interference. We highlight 3 advantages : i) the method lies in recording the free evolution of the interference fringes, neither spectrometer nor stabilisation system are required; ii) the value of CD is directly obtained thanks to an elegant formalism inferring the CD coefficient directly from the visibility of the two-photon interference; iii) this free-alignment method is definitively oriented towards real-world applications.









\section{Theoretical framework}

\begin{figure}[!ht]
    \centering
    \includegraphics[width=1\linewidth]{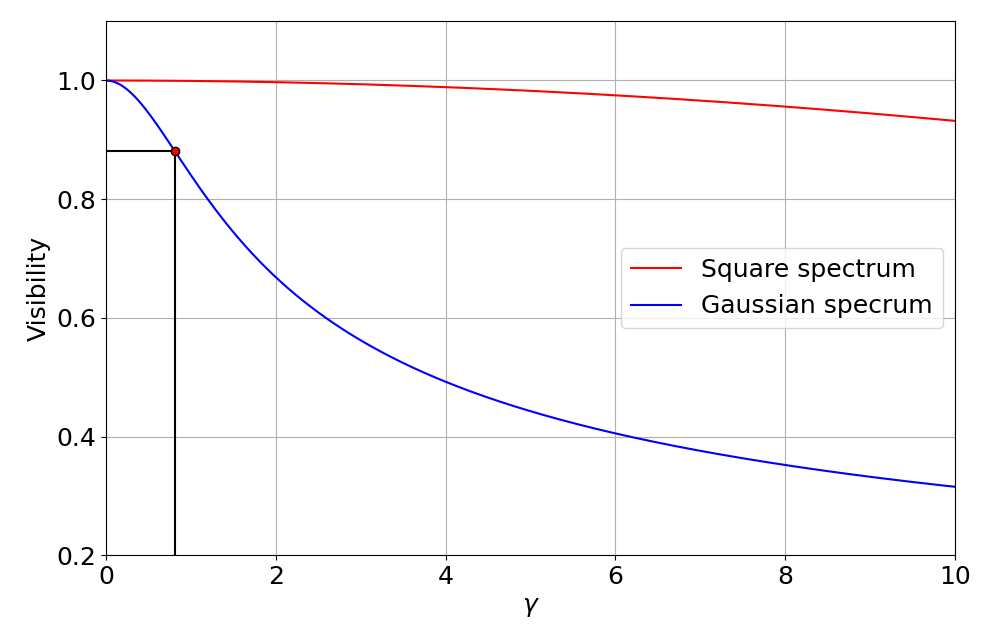}
    \caption{Visibility of the two-photon interference (V) as a function of the parameter $\gamma$, assuming Square/Gaussian functions represented by the red/blue curves, respectively. The red dot shows the inflexion point from Eq.~\ref{eq:second_derivative_Vis_Gauss}.}
    \label{fig:Visibility_theory}
\end{figure}

Frequency entangled photons generated via spontaneous parametric downconversion (SPDC) can be described by the 2-photon state:
\begin{equation}\label{eq:2photonstate}
    |\psi(t)\rangle \propto \iint d\omega_{s}d\omega_{i}   \alpha(\omega_s + \omega_i)\Gamma(\omega_s,\omega_i)
    \hat{a}^+_{s}\hat{a}^+_{i} |0\rangle,
\end{equation}
where subscripts s,i denote signal and idler photons, respectively. $\eta$,  $\alpha$, $\Gamma$ represent the strength of the non-linear interaction, the complex amplitude of the pump spectrum (approximated by a Dirac function in the continuous regime), and the spectral distribution of the photon pair, \textit{i.e} the phase matching function, respectively. The revelation of the entanglement carried by this bi-photon state lies in the two-photon interference through an appropriate unbalanced interferometer, often referred to a Franson configuration~\cite{franson_bell_1989, aktas_entanglement_2016}. It should be noted that in our case, we regard the two-photon state propagating within the interferometer as equivalent to an N00N state, achieved through post-selection during detection~\cite{Autebert, oser_high_quality_2020}. The visibility of the two-photon interference depends on the indistinguishability of the two paths in term of losses, polarization, spatial modes, as well as chromatic dispersion. More specifically, considering Taylor expansion of the wave-vector around the degeneracy wavelength, odd-order terms of the dispersion vanishes thanks to energy conservation~\cite{Riazi:19} associated with SPDC. The phase accumulated by the N00N state propagating within the interferometer can be expressed:
\begin{equation}\label{eq:phiNOON}
    \begin{split}
     \phi_{N00N} & = L\sum_{j=s,i}\sum_{n=0}^{\infty}\frac{\Delta\omega^n_j}{n!} \beta^{(n)}_j\\
    & = L\Big(2\beta^{(0)} + \beta^{(2)}\Delta\omega^2\Big) +O(\Delta\omega^4),   
    \end{split}
\end{equation}
where $\beta^{(n)}=\left.\frac{\partial k}{\partial \omega}\right|_{\omega_0}$, $\Delta\omega$ and $L$ stand as the detuning from the center frequency $\omega_0$ and  the length of the sample under test (SUT), respectively. The first term, $\beta^{(0)}$, represents a simple phase shift while the second term, $\beta^{(2)}$, is the definition of the CD. By performing a careful analysis of the evolution of the photon pair within the interferometer, using Eq.\ref{eq:2photonstate} and \ref{eq:phiNOON}, the visibility of the fringes reads :
\begin{equation}\label{eq:Vis_Beta}
 \begin{split}
    V & = \Bigg[\Big(\int_{-\infty}^{\infty}d\Delta\omega|\Gamma(\Delta\omega)|^2cos(\beta^{(2)}\Delta\omega^2)\Big)^2\\
    & +\Big(\int_{-\infty}^{\infty}d\Delta\omega|\Gamma(\Delta\omega)|^2sin(\beta^{(2)}\Delta\omega^2)\Big)^2\Bigg]^{1/2},
 \end{split}
\end{equation}
(see appendix A for details). The evolution of the visibility highly depends on the spectral distribution shape of the photon pairs, as shown in \figurename{\ref{fig:Visibility_theory}} where gaussian and square shapes are considered. In the case of a Gaussian distributed spectrum, \textit{i.e} $|\Gamma(\Delta\omega)|^2=\frac{1}{\sigma\sqrt{2\pi}}
e^{\frac{-1}{2}(\frac{\Delta\omega}{\sigma})^2}$, 
with spectral width $\sigma$, the visibility can be written under the analytic form:
\begin{equation}\label{eq:Vis_Gauss}
    V = \frac{1}{\sqrt[4]{\gamma^2+1}},
\end{equation}
with $\gamma=2\sigma^2\beta^{(2)}L$ represents an adimensional parameter carrying all the information on dispersion (sample length $L$, spectral bandwidth $\sigma$, and dispersion coefficient value $\beta^{(2)}$). This simple form indicates a straightforward relation between the visibility and the CD parameter.
In other words, when photon pair probes an object, information about the dispersive behavior of this object is imprinted on the two-photon state. This direct access to CD appears as an unique feature as temporal based CD measurements usually give access to the group delay before inferring the CD parameter.
It is important to note that the sensitivity to $\gamma$ depends on the spectral profile of the photon pairs, as also shown in \figurename{\ref{fig:Visibility_theory}}. The highest sensitivity is obtained when the leading coefficient of the tangent is maximum, namely the inflexion point. Considering only positive values of $\gamma$, the latter is located at the (only) inflexion of the curve where the second order derivative is null :
\begin{equation}\label{eq:second_derivative_Vis_Gauss}
    \frac{\partial^2 V}{\partial\gamma^2} = \frac{3\gamma^2 -2}{4(\gamma^2+1)^{9/4}} \Leftrightarrow \gamma = \sqrt{\frac{2}{3}},
\end{equation}
corresponding to a visibility of $V=(\frac{5}{3})^{-1/4}\approx 0.88$.\\
In summary, assuming a Gaussian spectral distribution, a direct access to the CD parameter is granted thanks to the precise knowledge of the visibility, which stands as a parameter easily accessible experimentally, without the need for spectrometer nor active stabilization system.

\section{Two-photon interference visibility estimation}
\label{section_visibility}

\begin{figure}[!ht]
    \centering
    \includegraphics[width=1\linewidth]{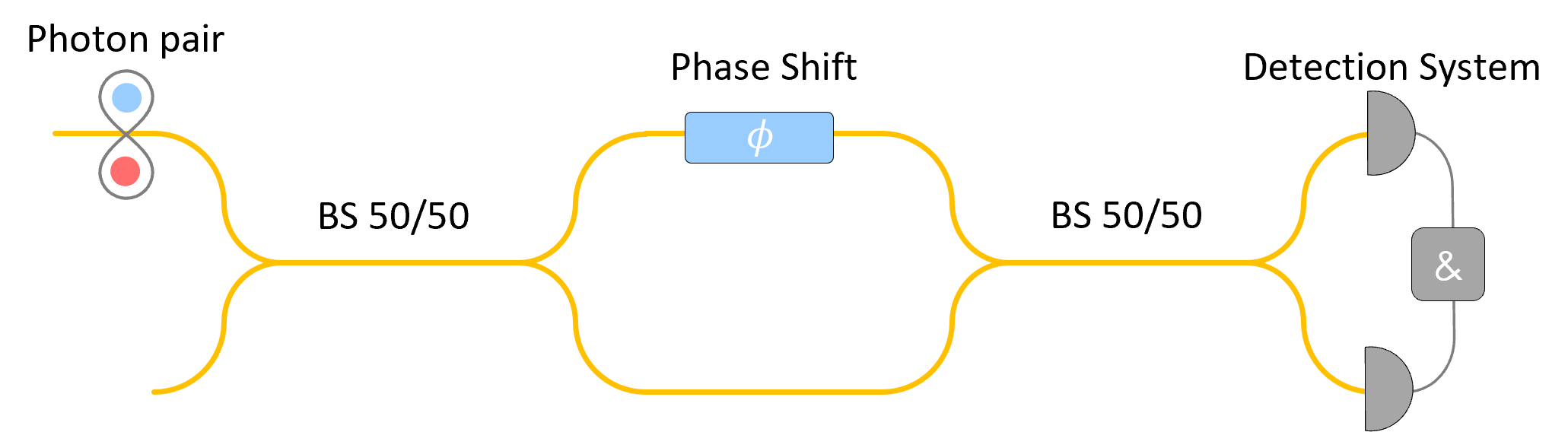}
    \caption{A photon pair experiences a Mach-Zenhder interferometer in which a phase shifter is introduced. At the output beam splitter both beams interfere creating interference patterns }
    \label{fig:diagram_scheme}
\end{figure}

\begin{figure*}[!ht]
    \centering
    \includegraphics[width=1\linewidth]{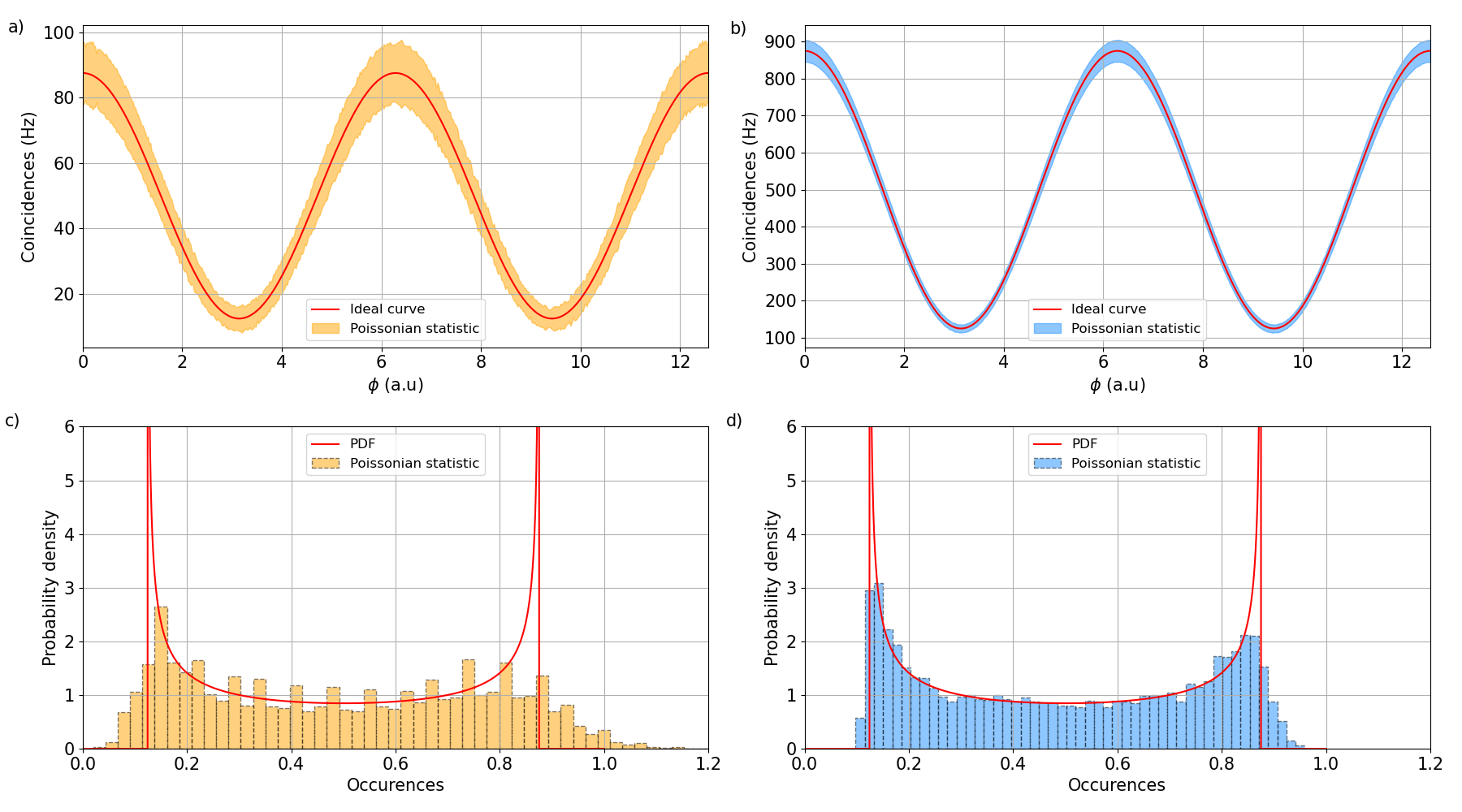}
    \caption{Top: Simulated Eq.~\ref{eq:Pc} with $V=0.75$ and $P_c(max) = 100$, (a)  and  $P_c(max) = 1000$ (b). Filled curves represent the statistical fluctuation due to the Poissonian distribution of the detection. Bottom: theoretical probability density function (red curve). Histograms are generated using Eq.~\ref{eq:Pc} for $N=10^4$ occurrences, with $\phi\in[0,2\pi]$ randomly distributed. The orange (c) and the blue (d) histograms refer to (a) and (b), respectively}
    \label{fig_pdf_theory}
\end{figure*}
In most of optical experiments, stabilisation of the interferometric system is of utmost importance in order to acquire precise and accurate measurements. However, many environmental parameters (e.g. temperature, pressure, vibration) impact the final result, requiring a fine control. In interferometric systems, both passive and active stabilisation are often considered. For passive stabilisation, interferometers are placed in an hermetic case with a thermal feedback. While this method is easy to implement, short- and long-term thermal drift cannot be managed in the same way, excluding high-degree of accuracy and long runs. On the other hand, active stabilisation, commonly based on an error signal generated from the output of the interferometer and fed back to a phase-compensating mechanism in one of its arms, is more robust against any kind of variations, at the price of more expensive and complex apparatus (e.g narrow-linewidth and stable laser, wavelength demultiplexer). 
Here, we show theoretically and experimentally an interferometric method based on the free-evolution of two-photon interference, avoiding the use of spectrometer and stabilization systems.\\
FIG.~\ref{fig:diagram_scheme} shows a diagram of the general principle of our method. The coincidence probability at the output of the interferometer is given by:
\begin{equation}\label{eq:Pc}
    P_c(\phi) \propto \frac{1}{2}\big[1+V cos(\phi + \phi_0)\big],
\end{equation}
where $V$ represents the visibility of the two-photon interference:
\begin{equation}\label{eq:V}
   V=\frac{C_{max}-C_{min}} {C_{max}+C_{min}},
\end{equation}
with $C_{max}$ and $C_{min}$ the maximum and minimum coincidence rates, respectively. 
The cumulative distribution function (CDF) is defined as the cumulative density of the variable $\phi$, $\Psi(x)=P_c(\phi) < P_c(x)$:
\begin{equation}\label{eq:CDF}
    \Psi(x) = \frac{asin(\frac{2}{V}(x-\frac{1}{2}))}{\pi} + \frac{1}{2}.
\end{equation}
This function monotonically increases and is related to the probability of the parameter $\phi$ to take a value below $x$. The probability density function (PDF), $\xi(x)$, which therefore sets the visibility likelihood for a distribution of coincidences, is defined as the derivative of the aforementioned function $\Psi(x)$:
\begin{equation}\label{eq:PDF}
    \xi(x) = \frac{2}{\pi V \sqrt{1+\frac{-4x^2 + 4x -1}{V^2}}}.
\end{equation}

In order to illustrate this formalism, we simulate the effect of poissonian statistic on the parameter estimation V and the PDF through Eq.~\ref{eq:PDF}.
As shown in \figurename{\ref{fig_pdf_theory}}, we set a two-photon interference visibility to 0.75 with, on average, 100 (a) and 1000 (b) photon-pairs. 
The red curves represent the theoretical prediction without finite statistic and the yellow and blue filled curves represent the spread due to the Poissonian distribution for 100 and 1000 photon-pairs, respectively.
The PDF functions, presented in \figurename{\ref{fig_pdf_theory}} (c,d), show that the probabilities are very close to the extreme values of 0.125 and 0.875 (corresponding to a visibility V$=$0.75) as most of the possible values spread over the extremes.
The Poissonian statistic of the detection induce modifications of the PDF: i) the edges spread over the two extreme values, ii) an asymmetry between these two extreme values arises. 
The first behavior naturally emerges from the Poissonian noise while the second one comes from the assymetry of the Poissonian distribution $\gamma = \lambda^{\frac{1}{2}}$, where $\lambda$ stands as the parameter of Poissonian distribution.
In other words, the relative error of the distribution is inversely proportional to the number of events. This model clearly shows the necessity of considering high coincidence rate ($>10^3$) to mitigate the impact of the statistic.  


Our approach based on the exploitation of the two-photon interference is reminiscent of the seminal violation of Bell's inequality applied to interferometric, or also called Franson, configuration in the 1980's~\cite{franson_bell_1989}. This interferometric technique used to measure quantum correlation induced by energy-time entanglement~\cite{aktas_entanglement_2016}, has opened the way to a wide range of applications such as non-local dispersion cancellation to ensure the security of Franson-based quantum key distribution protocols~\cite{scarani_security_2009, zhong_nonlocal_2013, zhang_unconditional_2014}. All cases require a precise knowledge of the visibility. Two strategies are primarily exploited. The first one relies on Eq.~\ref{eq:V}, where the extreme points $C_{max}$ and $C_{min}$ are inferred from the data analysis. On the other hand, the interference pattern can be fitted by Eq.~\ref{eq:Pc}. As the offset always fluctuates over the acquisition time, only the first oscillations are considered. None of the methods were convincingly enough as they consider only a part of the statistics. By contrast, we highlight the originality and the relevance of our approach which relies on the acquisition of long sets of data, each one being part of the PDF function. By the way, all the statistic contribute to the precise knowledge of the V parameter.

\section{Experimental results}

\begin{figure}
    \centering
    \includegraphics[width=1\linewidth]{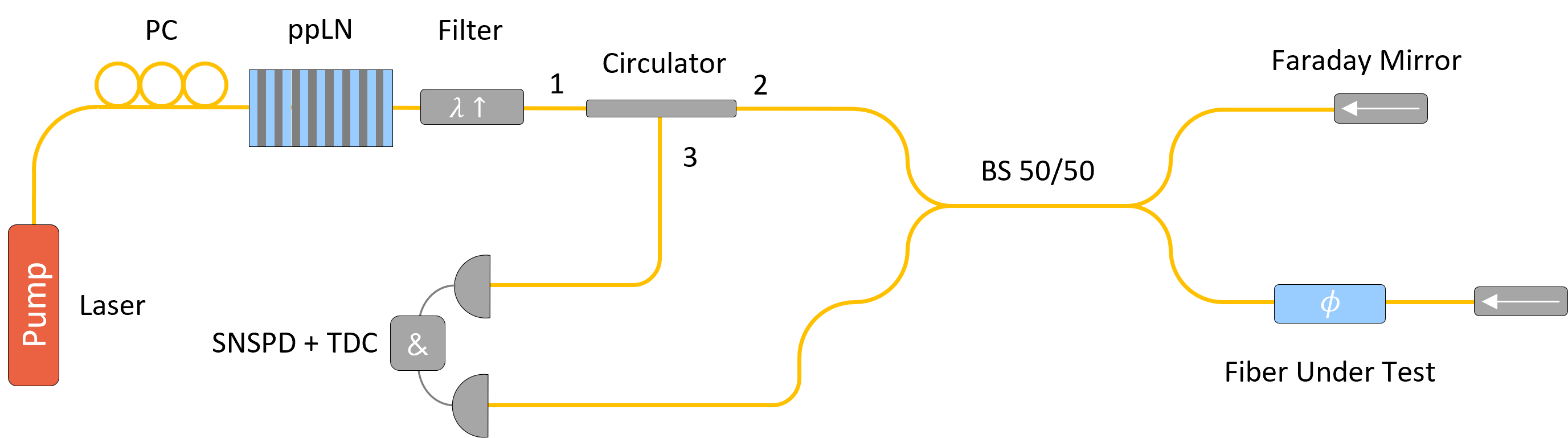}
    \caption{Experimental setup. A CW laser pumps a periodically poled lithium niobate waveguide (ppLN) from which correlated photon pairs are emitted via SPDC. These pairs are spectrally shaped with a  filter before passing trough a Michelson interferometer. Coincidence counts are registered using superconducting nanowire single photon detectors (SNSPD) and time digital converter (TDC). BS: 50/50 beam-splitter, PC: polarization controller}
    \label{fig:experimental_scheme}
\end{figure}

\begin{figure}[!ht]
    \centering
    \includegraphics[width=1\linewidth]{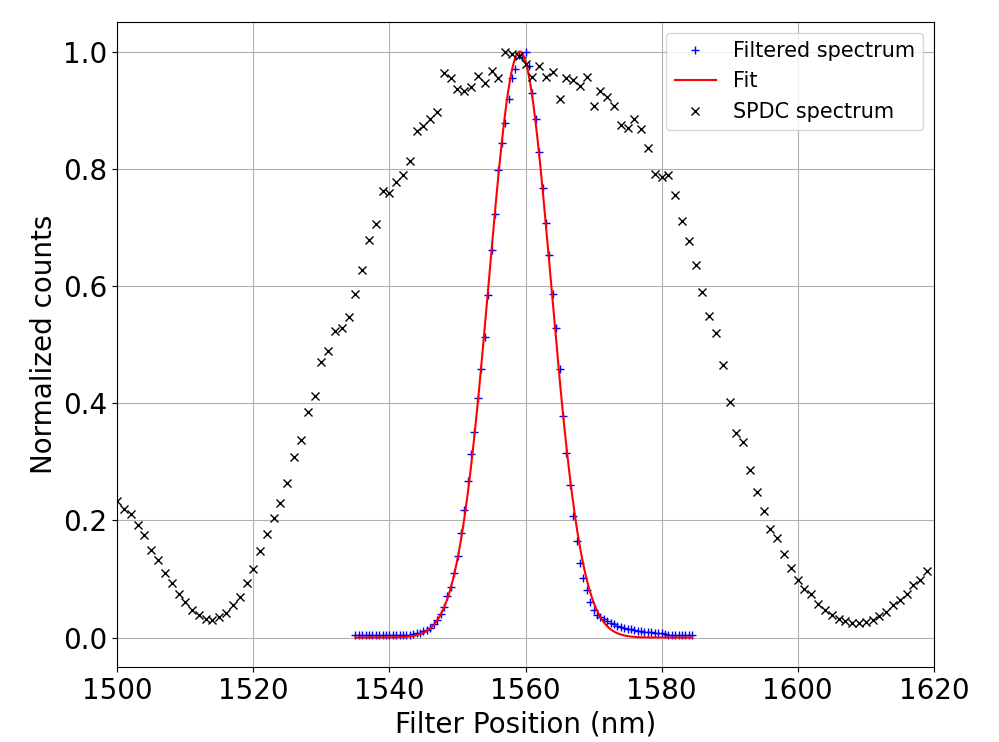}
    \caption{Spectrum of the photon pairs. The black dotted curve correspond to the natural spectrum of the pairs. The blue dotted curve is the spectrum measured after the band-pass filter. The red curve is a fit of the latter, from which we extract a width of $\sigma=4.57\,nm$. SPDC: pontaneous down-conversion}
    \label{fig:spectrum}
\end{figure}

A CW laser (Toptica TA Pro) at 780.23\,nm is sent in a polarisation controller followed by a pigtailed type-0 periodically poled lithium niobate (PPLN) waveguide in order to generate frequency correlated photon pairs, through spontaneous down-conversion (SPDC) process, symmetrically thanks a type-0 around the degenerated wavelength $\lambda_0 = 1560.46$\,nm, as shown in \figurename{\ref{fig:experimental_scheme}}. 
The phase matching condition of the non-linear interaction is adjusted to generate photon-pairs with a flat-top spectrum apart the degeneracy wavelength. 
Then the pairs are shaped thanks a filter to obtain a Gaussian spectrum centered around twice the pump laser wavelength (see \figurename{ \ref{fig:spectrum}}).
Photon pairs travel along a fiber-based interferometer where the sample under test (SUT) is a standard SMF-28 fiber.
Faraday mirrors are placed at the output of each arm, ensuring mode polarisation indistinguishability.
Finally, coincidences are recorded using superconducting nanowire single photon detectors (SNSPD) and a Time-to-Digital Converter (TDC).
As our method relies on the acquisition of the interferences operating in a free-running regime, a careful analysis has to be driven to ensure that the bandwidth of the detection system is higher than the phase drift.

\begin{figure*}[!ht]
    \centering
    \includegraphics[width=1\linewidth]{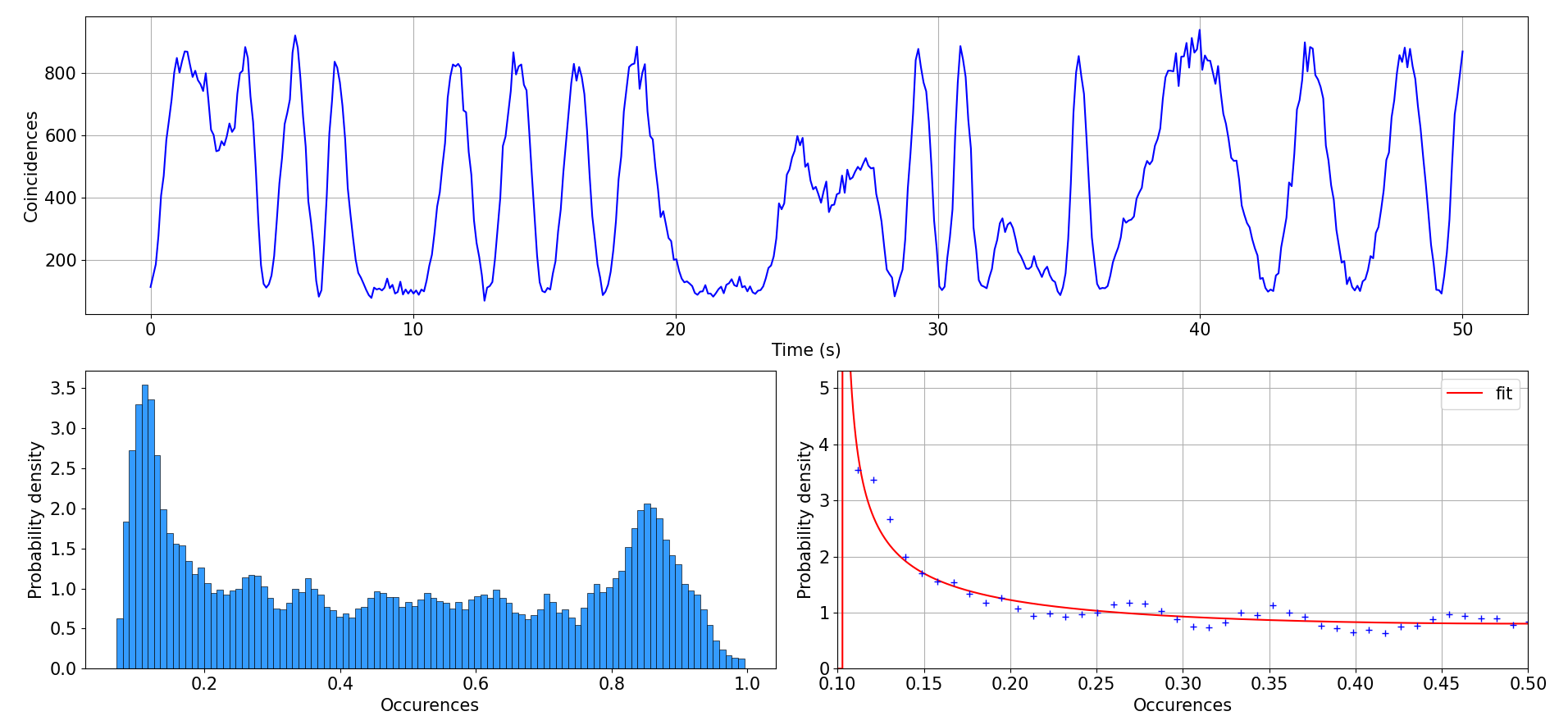}
    \caption{Top (a): Acquired coincidences (central peak of the Franson histogram) as a function of time. Bottom left (b): Extracted histogram from acquired data. Bottom right (c): fitted curve (red) of the left part of (b), using equation \ref{eq:PDF}.}
    \label{fig:experimental_pdf}
\end{figure*}

From the temperature-dependent refractive index of fused silica corresponding  to $\dfrac{\Delta n}{n} \approx 4.8.10^{-6}.K^{-1}.\Delta T$, with $\Delta T$ represents the temperature shift in K \cite{leviton_temperature-dependent_2008}, the sensitivity for a standard fiber can be extracted and is $0.2$\,K/m per fringes, which corresponds to a variation of 45\,mK in case of a 2.4\,m long fiber. By acquiring the temperature of the experiment room for 10 hours, the Fourier transform of the latter shows that typical variations have an amplitude of $\approx 35$\,mK and a frequency $f\in[0.5,2]$\,Hz, which is consistent with a 100\,ms acquisition time.

We set two different strategies for measuring the CD parameter. We extract the visibility: i) at a fixed operating point $\gamma$, corresponding to the inflexion point (see \figurename{\ref{fig_pdf_theory}} (a), and then Eq~\ref{eq:Vis_Gauss} is reverted, ii) for different values of $\gamma$ corresponding to different spectral bandwidth of the filter, fitted by Eq. \ref{eq:Vis_Gauss}.\\
For both approaches, a prerequisite has to be fulfilled to guarantee that the CD is the only factor impacting the visibility, ruling out other distinguishability criterion as differential losses, spatial, time and polarization modes. A calibration is fulfilled by measuring a 100\% visibility by considering narrow filtering, ensuring all the latter criterion are satisfied.
\\ 
\textit{Method based on the inflexion point}. Regarding \figurename{\ref{fig_pdf_theory}}, we need to make a trade-off concerning the pump power regime: a high number of coincidences ensuring that the statistics tends to the PDF, while maintaining a double-pair generation rate low enough to not deteriorate the visibility (typically $<$ 0.01 pairs/window of interest), due to accidental coincidences. The pump power is set to generate a maximum coincidence rate of $P_c \approx 10^4$\,Hz, while the FWHM of the Gaussian filter is set to $4.5\,nm$, giving a visibility close to 0.88, in accordance with the inflexion point of Eq.~\ref{eq:Vis_Gauss}. The raw spectrum out of the PPLN and the Gaussian-like filtered spectrum are shown in \figurename{\ref{fig:spectrum}}.

Typical two-photon fringes drifting in a free-running regime are shown in \figurename{\ref{fig:experimental_pdf}}(a). An histogram is built from a set of 500 points (50s) in \figurename{\ref{fig:experimental_pdf}}(b), and then is fitted thanks to Eq.~\ref{eq:PDF}. As expected, the PDF is shaped by the inherent Poissonian statistic. While the relative error induced by a Poissonian distribution is lower for high values, the absolute error is smaller for low values. In order to minimize the fitting error induced by the distribution, only the left part of the histogram is used (see \figurename{\ref{fig:experimental_pdf}}(c)), where the absolute error is the smallest. Then, Eq.~\ref{eq:Vis_Gauss} is reversed to extract the CD. We obtain after 200 measurements (\figurename{\ref{fig:GVD_experimental_full}}(a)), a CD parameter of $17,1(2)$ ps/(nm.km).


\begin{figure}[!h]
    \centering
    \includegraphics[width=0.8\linewidth]{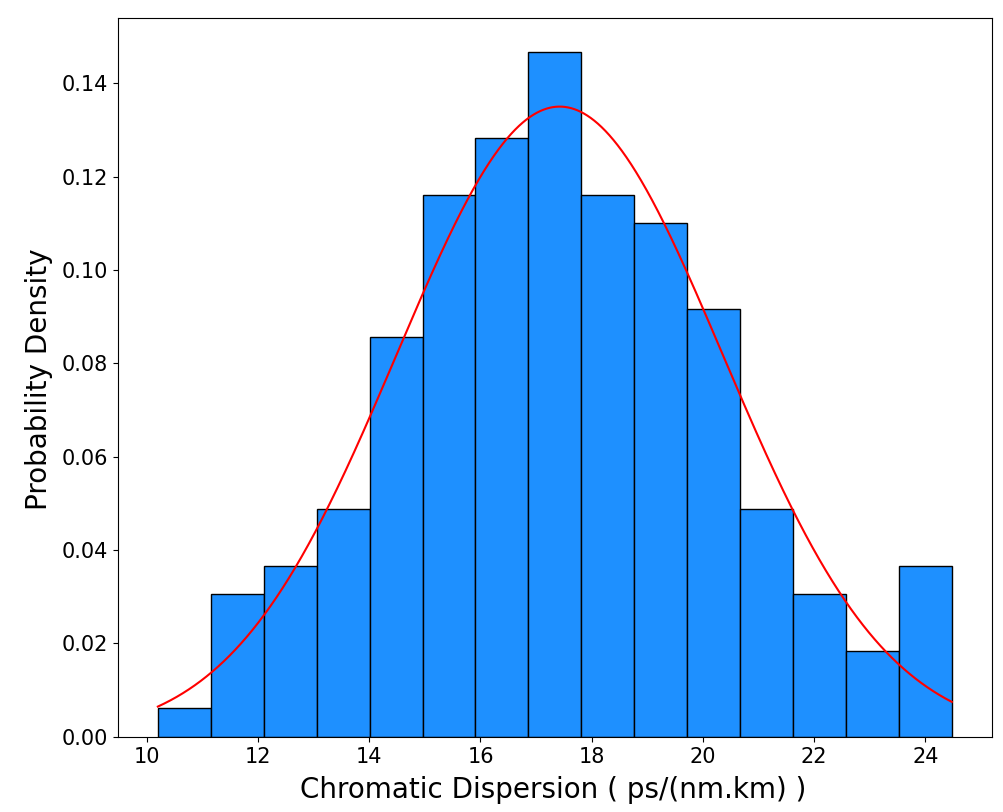}
    \caption{Chromatic dispersion measurement from 200 data samples (blue histogram) and its Gaussian fit (red). }
    \label{fig:GVD_experimental_full}
\end{figure}

\textit{Method based on multiple operating points.} The CD parameter is extracted by fitting Eq.~\ref{eq:Vis_Gauss}. Two scenarii are possible to tune the parameter $\gamma$ over the x-axis in \figurename{\ref{fig:Visibility_theory}}, either the length of the sample or the spectral bandwidth of the filter. This latter parameter is easily accessible as most of band-pass filter can be tunable, strengthening our vision of user-friendly demonstrator. The length of the sample shall be such that the visibility curve can be described with standard tunable filters, whose its spectral bandwidth ranges from $\sim$0.1~nm to $\approx$ 10~nm. According \figurename{\ref{fig:Visibility_theory}}, the length of the sample is set to 4.5\,m. As the previous method, a calibration is proceeded prior to any measurement ensuring 100\% visibility by considering narrow filtering. To keep the same poissonian statistic with respect to the different spectral bandwidth, the pump power increases as the bandwidth of the filter reduces so that the maximum number of coincidence remains the same $P_c \approx 10^4$\,Hz. The number of points acquired as well as the procedure of the visibility extraction remains identical to the first method, described in Sect.\ref{section_visibility}.

\begin{figure}[!h]
    \centering
    \includegraphics[width=0.8\linewidth]{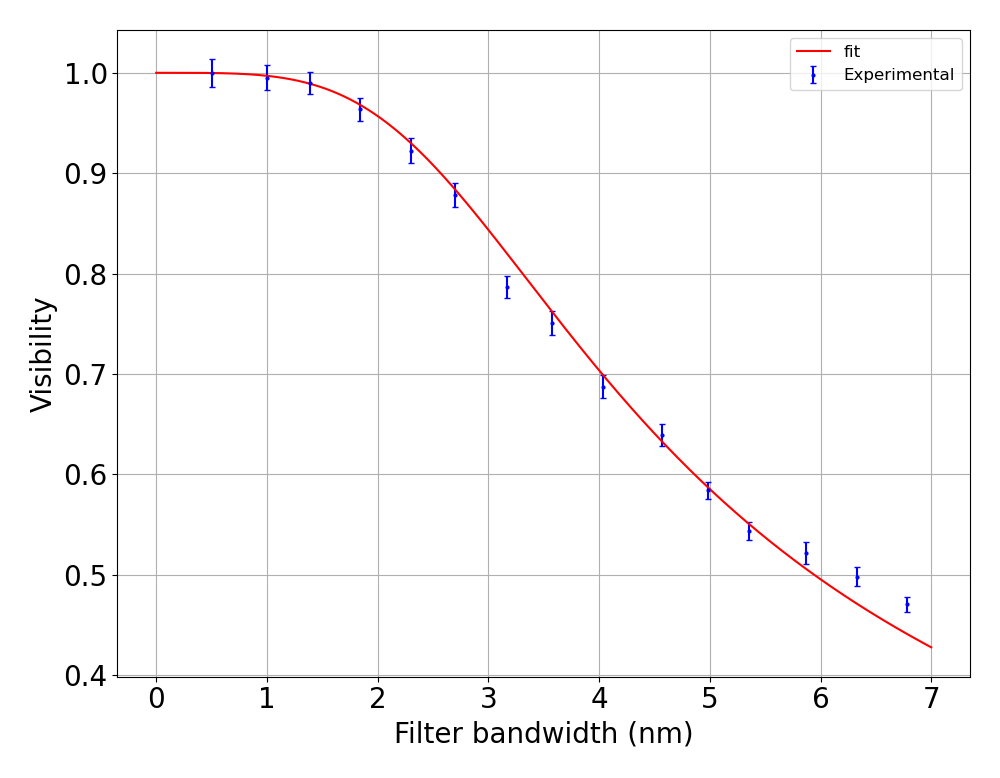}
    \caption{Extracted visibility as a function of the filter bandwidth. The number of measurements for each points is 200.}
    \label{fig:GVD_experimental_full}
\end{figure}

\figurename{\ref{fig:GVD_experimental_full}} shows the visibility values associated with each filter spectral width. The graph is fitted using Eq.~\ref{eq:Vis_Gauss} with only a free parameter being the CD parameter, leading to a CD value equal to 17.2(2)~ps/(km.nm).\\
The values of the CD parameter are 17.1(2)~ps/(km.nm) and 17.2(2)~ps/(km.nm) for both methods. Both approaches give results that are consistent each of them within 1\% of accuracy, and also with the manufacturer's data that predict a value $\leq$ 18~ps/(km.nm)\cite{Corning}. We would have expected that the first method would have given a better accuracy as it relies on the highest sensitive point, whereas the second method implements points with different sensitivities. This intuition has to be mitigated by a most favourable statistic because a higher number of runs for the second method has been performed, 3000 instead of 200 measurements.\\

We summarize in Table\,\ref{Tab:TabMetro} the main characteristics for different CD measurement methods, including both classical and quantum approaches. Even if the accuracy appear as a primary importance for a measurement technique, other criteria should be considered. Exhibiting a precision on the CD parameter of sub \% is requiring for non-linear effects which are sensitive to higher-order of dispersion $\beta_i$ (with $i\ge 2$), as four-wave mixing or modulation instability. But in most cases as telecommunication networks, sub \% accuracy is not necessary whereas friendly-user, practical and intuitive techniques are preferred, explaining the prevalence of CD apparatus based on temporal method in the industry. Our approach stands a trade-off among the different methods, allowing to exhibit a high accuracy while keeping a simple implementation with a direct access to the dispersion thanks to unique features of quantum photonics. We emphasize the novelty of our approach based on free running acquisition combining the simplicity of temporal methods with the accuracy of spectral methods, leading the potential to compete with well-established conventional techniques. Beyond the measurement of CD, this method opens a new application path, that of probing the dispersive properties of materials, in the same way as distance measurement for optical coherence tomography  or transmission/absorption for spectroscopy. 

\begin{table*}[t]
\begin{center}
\begin{tabular}{|c||c|c|c|c|}
\hline
Method & CD accuracy & Sample length & Active stabilization & Dedicated equipment \\\hline
Phase shift~\cite{baker_chromatic-dispersion_2014} & 0.1\% & km & No & Standard \\\hline
Time of flight~\cite{sucbei_moon_reflectometric_2009} & 3\% & m & No & Standard \\\hline
WLI~\cite{grosz_measurement_2014} & 1.5\% & [cm; m] & Yes & Spectrometer + balanced interferometer \\\hline
QWLI~\cite{kaiser_quantum_2018} & 0.02\% & [cm; m] & Yes & Spectrometer\\\hline
Our work & 1\% & [cm; m] & No & Standard \\\hline
\end{tabular}
\end{center}
\caption{\label{Tab:TabMetro}Comparison between different CD measurement technique. WLI and QWLI denotes white-light interferometry and quantum white-light interferometry, respectively}
\end{table*}%

\section{Discussion}
We have implemented a new technique for measuring the CD through a direct relation between the visibility of the two-photon interferences and the CD parameter, based on the assumption of a gaussian profile of the two-photon states. This model goes beyond the scope of probing dispersive properties in material, and could be useful for a wide variety of applications requiring an accurate knowledge of the visibility as for ensuring security for quantum key distribution protocols. The key point of our technique gathers the simplicity of implementation of temporal technique associated to high accuracy of spectral ones without the requirement of active stabilization. We have shown that our technique is compliant with two complementary variations resulting to a similar 1\% of accuracy. This work represents a step towards a realistic and a friendly-user quantum enhanced demonstrator, outperforming classical counterparts (accuracy, reproducibility and immunity to environment) demonstrated so far for the last 20 years. Our strategy focuses on a specific use-case in this present work, but has to be included in a more general framework, within the acceleration of quantum technologies, of emerging from the laboratory new generation of quantum-enhanced sensors.

\section*{Acknowledgment}
This work has been conducted within the framework of the project OPTIMAL granted by the European Union by means of the Fond Européen de développement regional (FEDER). The authors also acknowledge financial support from the Agence Nationale de la Recherche (ANR) through the projects METROPOLIS, the CNRS through its program “Mission interdisciplinairité” under project labeled FINDER, and the French government through its Investments for the Future programme under the Université Côte d'Azur UCA-JEDI  project  (Quantum@UCA)  managed  by  the  ANR (ANR-15-IDEX-01).

\section*{Author information}

M.R. established the theory, R.D. and G.S. performed the experiments under the supervision of A.M.\\
All the authors contributed to the the paper.

\section*{Competing interests}
The authors declare that there are no competing interests.

\section*{Data Availability}
Data are available from the authors on reasonable request.

\onecolumngrid

\section{Supplementary Data}

\subsection{Evolution of input mode \textit{a} through Michelson interferometer}

 The creation operator $a^{\dagger}$ at the input of the interferometer evolve as $a^{\dagger} \rightarrow \frac{1}{\sqrt{2}}(a^{\dagger} + b^{\dagger})$ at the first beamsplitter, $a^{\dagger}$ and $b^{\dagger}$ representing the lower and upper path of the interferometer respectively. The operators $c^{\dagger}$ and $d^{\dagger}$ are the two output modes of the interferometer, i.e. at the output of the second beamsplitter. The  phase  that each photon is accumulating when passing either trough the upper or lower arm, is noted $\Psi_{a,b}(\omega)$. This function can be developed around a central frequency $\omega_0 = \frac{\omega_{pump}}{2}$ : 

\begin{equation}
 \Psi_i(\omega) = L_i\sum\frac{\partial^n k_i}{\partial\omega^n}\frac{(\omega_0-\omega)^n}{n!} = L_i\sum\beta_i^{(n)}\frac{\Omega^n}{n!}, 
\end{equation}

Where L is the length of the chosen arm. $\beta^{(0)}, \beta^{(1)}, \beta^{(2)}$ represents a simple phase shift, the inverse of group velocity and the group velocity dispersion, respectively. The evolution inside the interferometer is given by :

\begin{align}
a_{in}^\dagger &\rightarrow \frac{1}{\sqrt{2}}( a^\dagger + b^\dagger) \\
 &\rightarrow \frac{1}{\sqrt{2}}( e^{i\Psi_a(\omega)}a^\dagger + e^{i\Psi_b(\omega)} b^\dagger) \\
&\rightarrow  \frac{1}{2}(e^{i\Psi_a(\omega)} (c^\dagger + d^\dagger)+ e^{i\Psi_b(\omega)} (c^\dagger - d^\dagger)) \\
&=  \frac{1}{2}\left[ (e^{i\Psi_a(\omega)} + e^{i\Psi_b(\omega)}) c^\dagger + (e^{i\Psi_a(\omega)} - e^{i\Psi_b(\omega)}) d^\dagger \right] \\
\end{align}

\subsection{2-photon state through the interferometer}

Photon pairs are created by a down-conversion source e.g. type 0 periodically poled lithium niobate crystal. These pairs are spectrally filtered and enter the interferometer via input mode a. They can be described by the 2-photon state:
\begin{equation}
    |\psi_{in}\rangle = \eta \iint d\omega_{s,i}  \alpha(\omega_s + \omega_i)\Gamma(\omega_s,\omega_i)
    \hat{a}^+_{s,i} |0\rangle
\end{equation}

where subscripts s,i denote signal/idler photons.$\alpha$, $\Gamma$ represent  the complex amplitude of the pump spectrum (approximated by a Dirac function in the continuous regime) and the spectral distribution of the photon pair, \textit{i.e} the phase matching function, respectively. In the following lines, the evolution through the interferometer is regarded :

\begin{align}
 |\psi_{in}\rangle & = \iint d\omega_{s,i}\Gamma(\omega_s,\omega_i)\hat{a}^+_{s,i} |0\rangle \\
&=\frac{1}{4} \iint d\omega_{s,i} \Gamma(\omega_s,\omega_i) \left[ (e^{i\Psi_a(\omega_s)} + e^{i\Psi_b(\omega_s)}) c^\dagger_{s} + (e^{i\Psi_a(\omega_s)} - e^{i\Psi_b(\omega_s)}) d^\dagger_s \right] \\
&\qquad\qquad\qquad\qquad\quad \times \left[ (e^{i\Psi_a(\omega_i)} + e^{i\Psi_b(\omega_i)}) c^\dagger_i + (e^{i\Psi_a(\omega_i)} - e^{i\Psi_b(\omega_i)}) d^\dagger_i \right] |0\rangle\\ 
 |\psi_{out}\rangle & = \frac{1}{4} \iint d\omega_{s,i} \Gamma(\omega_s,\omega_i)  \left[ (e^{i\Psi_a(\omega_s)} + e^{i\Psi_b(\omega_s)})(e^{i\Psi_a(\omega_i)} + e^{i\Psi_b(\omega_i)}) c^\dagger_s c^\dagger_i  \right.\\
& \qquad \qquad\qquad\qquad\quad 
+(e^{i\Psi_a(\omega_s)} - e^{i\Psi_b(\omega_s)})(e^{i\Psi_a(\omega_i)} - e^{i\Psi_b(\omega_i)}) d^\dagger_s d^\dagger_i  \\
& \qquad \qquad\qquad\qquad\quad 
+(e^{i\Psi_a(\omega_s)} + e^{i\Psi_b(\omega_s)})(e^{i\Psi_a(\omega_i)} - e^{i\Psi_b(\omega_i)}) c^\dagger_s d^\dagger_i  \\
& \qquad \qquad\qquad\qquad\quad 
\left.+ (e^{i\Psi_a(\omega_s)} - e^{i\Psi_b(\omega_s)})(e^{i\Psi_a(\omega_i)} + e^{i\Psi_b(\omega_i)}) d^\dagger_s c^\dagger_i \right]  |0\rangle
\end{align}

\subsection{Coincidence Probability}

The coincidence probability $P_c$ is given by the projection of the output-state $ |\psi_{out}\rangle$ onto the state $c^\dagger_{\omega'} d^\dagger_{\omega''} |0\rangle = |{\omega_c \omega'_d}\rangle$, that corresponds to detecting one photon at frequency $\omega$ in output-mode c and one photon at frequency $\omega'$ in output-mode d. Assuming a photo detector measuring photon absorption without distinguishing between different frequency components, the coincidence probability is equivalent to :

\begin{align}
P_c &= \iint d\omega d\omega'  \left| \langle {\omega_c \omega'_d} |{\Phi_{out}\rangle} \right|^2 
\label{eqn_Pc_1}
\end{align}
where the integration is over the monochromatic modes that enter the detectors. Since $[c_{\omega}, c^\dagger_{\omega'}] = \delta(\omega - \omega') $ and due to the symmetry of $G=G(\omega_s,\omega_i)=G(\omega_i,\omega_s)$, equation(15)  simplifies to :

\begin{align}
P_c &= \frac{1}{8} \iint d\omega d\omega' |\Gamma(\omega,\omega')|^2 \left| (e^{i\Psi_a(\omega)} + e^{i\Psi_b(\omega)})(e^{i\Psi_a(\omega')} - e^{i\Psi_b(\omega')}) \right|^2 \quad 
\label{eqn_Pc_2}
\end{align}

by posing $\omega = \omega_0 + \Omega $ and $\omega' = \omega_0 -\Omega$, rearranging the phase functions $\Psi_{a,b}(\omega)$ to form a total phase term , equation (16) can be written as :    

\begin{align}
P_c &= \frac{1}{8} \int d\Omega |\Gamma(\Omega)|^2 \left| (1+ e^{i\Psi(\Omega)})(1 - e^{i\Psi(-\Omega)}) \right|^2 \\
&= \frac{1}{8} \int^\infty_{-\infty} d\Omega |\Gamma(\Omega)|^2  (1+ e^{i\Psi(\Omega)})(1+ e^{-i\Psi(\Omega)})(1 - e^{i\Psi(-\Omega)})(1 - e^{-i\Psi(-\Omega)})  \\
&=\frac{1}{2} \int^\infty_{-\infty} d\Omega |\Gamma(\Omega)|^2 \left[1+ \cos(\Psi(\Omega))\right] \left[1 - \cos(\Psi(-\Omega)) \right] \\
&=\frac{1}{2} \int^\infty_{-\infty} d\Omega |\Gamma(\Omega)|^2 \left[1+ \cos(\Psi(\Omega)) - \cos(\Psi(-\Omega))  - \cos(\Psi(\Omega)) \cos(\Psi(-\Omega))  \right]
\end{align}

The term $\cos(\Psi(\Omega)) - \cos(\Psi(-\Omega))$ is an odd function, regardless the exact form of $\Psi(\Omega)$ while $|\Gamma(\Omega)|^2$ is a even (symmetric) function. Therefore,  $|G'(\Omega)|^2[(\cos(\Psi(+\Omega)) - \cos(\Psi(-\Omega)]$ is odd as well and the integral over the whole frequency-space equals zero. The term $\cos(\Psi(\Omega)) \cos(\Psi(-\Omega)) $ in contrast is even and thus contributes to $P_c$. It can be rewritten using product-to-sum identities for trigonometric functions as $\cos(\Psi(\Omega)) \cos(\Psi(-\Omega)) =\frac{1}{2}{\cos[\Psi(\Omega) +\Psi(-\Omega) ]} +\frac{1}{2}{\cos[\Psi(\Omega) -\Psi(-\Omega) ]}  $ . This separates the even ($\frac{\Psi(\Omega) +\Psi(-\Omega)}{2}$)  and odd ($\frac{\Psi(\Omega) -\Psi(-\Omega)}{2}$)) parts of $\Psi(\Omega)$. equation (20) then becomes :

\begin{align}
P_c &=\frac{1}{4} \int d\Omega |\Gamma(\Omega)|^2 \left\lbrace 2 -{\cos[\Psi(\Omega) +\Psi(-\Omega) ]} -{\cos[\Psi(\Omega) -\Psi(-\Omega) ]}  \right\rbrace   \\
&= \frac{1}{4} \left\lbrace 2 - \int d\Omega |\Gamma(\Omega)|^2 \cos[\Psi(\Omega) +\Psi(-\Omega)] - \int d\Omega |\Gamma(\Omega)|^2 \cos[\Psi(\Omega) -\Psi(-\Omega) ]  \right\rbrace 
\end{align}

\subsection{Evaluation and interpretation}

In order to understand what each term in equation (22) is describing, we primarily neglect the second and higher order terms of the dispersion :

\begin{align}
P_c &= \frac{1}{4} \left\lbrace 2 - \int d\Omega |\Gamma(\Omega)|^2 \cos[2\beta^{(0)}L] - \int d\Omega |\Gamma(\Omega)|^2 \cos[2 \Omega\beta^{(1)}L ]  \right\rbrace \\
&= \frac{1}{4} \left\lbrace 2 - \cos(2\beta^{(0)}L)- \hat{G'}\star \hat{G}(2 \beta^{1}L) \right\rbrace
\end{align}
where $\hat{G(t)}= \int^\infty_{-\infty} d\Omega \Gamma(\Omega) \cos[\Omega t  ] = \int^\infty_{-\infty} d\Omega \Gamma(\Omega) e^{i\Omega t }$ stands for the Fourier transform of $\Gamma(\Omega)$ and $\hat{G}\star \hat{G}(t)=\int^\infty_{-\infty} ds \hat{G(s)} \hat{G^{*}(s-t)}$ for the autocorrelation of $\hat{G}$.

This is the well known two-photon-state-interferogram, where the first term originates from all the possible distinguishable paths, the second term comes from two photons (NOON-state, with N=2)) traveling along the same path, resulting in a Franson-type oscillation and the third term corresponds to the interference of two identical monochromatic modes , equivalent to the Hong-ou-Mandel (HOM) effect. \\

Including now second and third-order dispersion in our model, integration of equation (22) is not trivial anymore. The The Term corresponding to the Franson-Type-Oscillation becomes : 

\begin{align}
F&=\int d\Omega |\Gamma(\Omega)|^2 \cos[\Psi(\Omega) +\Psi(-\Omega)]  \\
&= \int d\Omega |\Gamma(\Omega)|^2 \cos[2\beta^{(0)}L + \beta^{(2)}\Omega^{2} L]
\\
&= \int d\Omega |\Gamma(\Omega)|^2 \left[ \cos(2\beta^{(0)}L)\cos(\beta^{(2)} \Omega^{2} L) -  \sin(2\beta^{(0)}L)\sin(\beta^{(2)} \Omega^{2}) L \right]
\\
&= \cos(2\beta^{(0)}L) \cdot \int d\Omega |\Gamma(\Omega)|^2 \cos(\beta^{(2)} \Omega^{2} L) -  \sin(2\beta^{(0)}L) \cdot \int d\Omega |\Gamma(\Omega)|^2 \sin(\beta^{(2)} \Omega^{2} L) \\
&= V_D \cdot \cos(2\beta^{(0)}L + \psi_D) \quad ,
\end{align}
where 
\begin{equation}
\psi_D= \tan^{-1} \left(\frac{\int d\Omega |\Gamma(\Omega)|^2 \sin(\beta^{(2)} \Omega^{2} L)}{\int d\Omega |\Gamma(\Omega)|^2 \cos(\beta^{(2)} \Omega^{2} L)} \right)
\label{eqn_phase}
\end{equation}
and
\begin{equation}
V_D=\sqrt{\left( \int d\Omega |\Gamma(\Omega)|^2 \cos(\beta^{(2)} \Omega^{2} L)\right)^2 + \left( \int d\Omega |\Gamma(\Omega)|^2 \sin(\beta^{(2)} \Omega^{2} L)\right)^2}
\label{eqn_vis}
\end{equation}
are the phase and the visibility of the oscillation. Using the Cauchy–Schwarz inequality one can find that $V_D \leq 1$, but the exact value is highly depending on the spectrum $\Gamma(\Omega)$. \\

Assuming a Gaussian distributed spectrum of bandwidth $\sigma$, the visibility and the phase of the oscillation can be calculated as a function of $\gamma=2\sigma^2\beta^{(2)}L$, a constant that combines the spectral width and chromatic dispersion :

\begin{equation}
\int d\Omega |\Gamma(\Omega)|^2 \cos(\beta^{(2)} \Omega^{2} L)=\sqrt{\frac{1 +\sqrt{1+\gamma^2}}{2 +2\gamma^2}}
\end{equation}
\begin{equation}
\int d\Omega |\Gamma(\Omega)|^2 \sin(\beta^{(2)} \Omega^{2} L)= \frac{\gamma}{\sqrt{2(1+\gamma^2)(1 + \sqrt{1+\gamma^2}}}
\end{equation}

Equation (31) thus become :

\begin{equation}
V_D(d)=\frac{1}{\sqrt[4]{\gamma^2 +1}}
\end{equation}

\bibliography{bib}

\begin{thebibliography}{30}%
\makeatletter
\providecommand \@ifxundefined [1]{%
 \@ifx{#1\undefined}
}%
\providecommand \@ifnum [1]{%
 \ifnum #1\expandafter \@firstoftwo
 \else \expandafter \@secondoftwo
 \fi
}%
\providecommand \@ifx [1]{%
 \ifx #1\expandafter \@firstoftwo
 \else \expandafter \@secondoftwo
 \fi
}%
\providecommand \natexlab [1]{#1}%
\providecommand \enquote  [1]{``#1''}%
\providecommand \bibnamefont  [1]{#1}%
\providecommand \bibfnamefont [1]{#1}%
\providecommand \citenamefont [1]{#1}%
\providecommand \href@noop [0]{\@secondoftwo}%
\providecommand \href [0]{\begingroup \@sanitize@url \@href}%
\providecommand \@href[1]{\@@startlink{#1}\@@href}%
\providecommand \@@href[1]{\endgroup#1\@@endlink}%
\providecommand \@sanitize@url [0]{\catcode `\\12\catcode `\$12\catcode
  `\&12\catcode `\#12\catcode `\^12\catcode `\_12\catcode `\%12\relax}%
\providecommand \@@startlink[1]{}%
\providecommand \@@endlink[0]{}%
\providecommand \url  [0]{\begingroup\@sanitize@url \@url }%
\providecommand \@url [1]{\endgroup\@href {#1}{\urlprefix }}%
\providecommand \urlprefix  [0]{URL }%
\providecommand \Eprint [0]{\href }%
\providecommand \doibase [0]{https://doi.org/}%
\providecommand \selectlanguage [0]{\@gobble}%
\providecommand \bibinfo  [0]{\@secondoftwo}%
\providecommand \bibfield  [0]{\@secondoftwo}%
\providecommand \translation [1]{[#1]}%
\providecommand \BibitemOpen [0]{}%
\providecommand \bibitemStop [0]{}%
\providecommand \bibitemNoStop [0]{.\EOS\space}%
\providecommand \EOS [0]{\spacefactor3000\relax}%
\providecommand \BibitemShut  [1]{\csname bibitem#1\endcsname}%
\let\auto@bib@innerbib\@empty
\bibitem [{\citenamefont {Wei}\ \emph {et~al.}(2022)\citenamefont {Wei},
  \citenamefont {Jing}, \citenamefont {Zhang}, \citenamefont {Liao},
  \citenamefont {Yuan}, \citenamefont {Fan}, \citenamefont {Lyu}, \citenamefont
  {Zhou}, \citenamefont {Wang}, \citenamefont {Deng}, \citenamefont {Song},
  \citenamefont {Oblak}, \citenamefont {Guo},\ and\ \citenamefont
  {Zhou}}]{Wei22}%
  \BibitemOpen
  \bibfield  {author} {\bibinfo {author} {\bibfnamefont {S.-H.}\ \bibnamefont
  {Wei}}, \bibinfo {author} {\bibfnamefont {B.}~\bibnamefont {Jing}}, \bibinfo
  {author} {\bibfnamefont {X.-Y.}\ \bibnamefont {Zhang}}, \bibinfo {author}
  {\bibfnamefont {J.-Y.}\ \bibnamefont {Liao}}, \bibinfo {author}
  {\bibfnamefont {C.-Z.}\ \bibnamefont {Yuan}}, \bibinfo {author}
  {\bibfnamefont {B.-Y.}\ \bibnamefont {Fan}}, \bibinfo {author} {\bibfnamefont
  {C.}~\bibnamefont {Lyu}}, \bibinfo {author} {\bibfnamefont {D.-L.}\
  \bibnamefont {Zhou}}, \bibinfo {author} {\bibfnamefont {Y.}~\bibnamefont
  {Wang}}, \bibinfo {author} {\bibfnamefont {G.-W.}\ \bibnamefont {Deng}},
  \bibinfo {author} {\bibfnamefont {H.-Z.}\ \bibnamefont {Song}}, \bibinfo
  {author} {\bibfnamefont {D.}~\bibnamefont {Oblak}}, \bibinfo {author}
  {\bibfnamefont {G.-C.}\ \bibnamefont {Guo}},\ and\ \bibinfo {author}
  {\bibfnamefont {Q.}~\bibnamefont {Zhou}},\ }\bibfield  {title} {\bibinfo
  {title} {Towards real-world quantum networks: A review},\ }\href@noop {}
  {\bibfield  {journal} {\bibinfo  {journal} {Laser \& Photonics Reviews}\
  }\textbf {\bibinfo {volume} {16}},\ \bibinfo {pages} {2100219} (\bibinfo
  {year} {2022})}\BibitemShut {NoStop}%
\bibitem [{\citenamefont {Flamini}\ \emph {et~al.}(2019)\citenamefont
  {Flamini}, \citenamefont {Spagnolo},\ and\ \citenamefont
  {Sciarrino}}]{Fla19}%
  \BibitemOpen
  \bibfield  {author} {\bibinfo {author} {\bibfnamefont {F.}~\bibnamefont
  {Flamini}}, \bibinfo {author} {\bibfnamefont {N.}~\bibnamefont {Spagnolo}},\
  and\ \bibinfo {author} {\bibfnamefont {F.}~\bibnamefont {Sciarrino}},\
  }\bibfield  {title} {\bibinfo {title} {Photonic quantum information
  processing: a review},\ }\href {https://doi.org/10.1088/1361-6633/aad5b2}
  {\bibfield  {journal} {\bibinfo  {journal} {Reports on Progress in Physics}\
  }\textbf {\bibinfo {volume} {82}},\ \bibinfo {pages} {016001} (\bibinfo
  {year} {2019})}\BibitemShut {NoStop}%
\bibitem [{\citenamefont {Schirhagl}\ \emph {et~al.}(2014)\citenamefont
  {Schirhagl}, \citenamefont {Chang}, \citenamefont {Loretz},\ and\
  \citenamefont {Degen}}]{NV}%
  \BibitemOpen
  \bibfield  {author} {\bibinfo {author} {\bibfnamefont {R.}~\bibnamefont
  {Schirhagl}}, \bibinfo {author} {\bibfnamefont {K.}~\bibnamefont {Chang}},
  \bibinfo {author} {\bibfnamefont {M.}~\bibnamefont {Loretz}},\ and\ \bibinfo
  {author} {\bibfnamefont {C.~L.}\ \bibnamefont {Degen}},\ }\bibfield  {title}
  {\bibinfo {title} {Nitrogen-vacancy centers in diamond: Nanoscale sensors for
  physics and biology},\ }\href
  {https://doi.org/10.1146/annurev-physchem-040513-103659} {\bibfield
  {journal} {\bibinfo  {journal} {Annual Review of Physical Chemistry}\
  }\textbf {\bibinfo {volume} {65}},\ \bibinfo {pages} {83} (\bibinfo {year}
  {2014})}\BibitemShut {NoStop}%
\bibitem [{\citenamefont {Geiger}\ \emph {et~al.}(2020)\citenamefont {Geiger},
  \citenamefont {Landragin}, \citenamefont {Merlet},\ and\ \citenamefont
  {Pereira Dos~Santos}}]{geiger}%
  \BibitemOpen
  \bibfield  {author} {\bibinfo {author} {\bibfnamefont {R.}~\bibnamefont
  {Geiger}}, \bibinfo {author} {\bibfnamefont {A.}~\bibnamefont {Landragin}},
  \bibinfo {author} {\bibfnamefont {S.}~\bibnamefont {Merlet}},\ and\ \bibinfo
  {author} {\bibfnamefont {F.}~\bibnamefont {Pereira Dos~Santos}},\ }\bibfield
  {title} {\bibinfo {title} {High-accuracy inertial measurements with cold-atom
  sensors},\ }\href {https://doi.org/10.1116/5.0009093} {\bibfield  {journal}
  {\bibinfo  {journal} {AVS Quantum Science}\ }\textbf {\bibinfo {volume}
  {2}},\ \bibinfo {pages} {024702} (\bibinfo {year} {2020})}\BibitemShut
  {NoStop}%
\bibitem [{\citenamefont {Taylor}\ \emph {et~al.}(2008)\citenamefont {Taylor},
  \citenamefont {Cappellaro}, \citenamefont {Childress}, \citenamefont {Jiang},
  \citenamefont {Budker}, \citenamefont {Hemmer}, \citenamefont {Yacoby},
  \citenamefont {Walsworth},\ and\ \citenamefont
  {Lukin}}]{taylor_high-sensitivity_2008}%
  \BibitemOpen
  \bibfield  {author} {\bibinfo {author} {\bibfnamefont {J.~M.}\ \bibnamefont
  {Taylor}}, \bibinfo {author} {\bibfnamefont {P.}~\bibnamefont {Cappellaro}},
  \bibinfo {author} {\bibfnamefont {L.}~\bibnamefont {Childress}}, \bibinfo
  {author} {\bibfnamefont {L.}~\bibnamefont {Jiang}}, \bibinfo {author}
  {\bibfnamefont {D.}~\bibnamefont {Budker}}, \bibinfo {author} {\bibfnamefont
  {P.~R.}\ \bibnamefont {Hemmer}}, \bibinfo {author} {\bibfnamefont
  {A.}~\bibnamefont {Yacoby}}, \bibinfo {author} {\bibfnamefont
  {R.}~\bibnamefont {Walsworth}},\ and\ \bibinfo {author} {\bibfnamefont
  {M.~D.}\ \bibnamefont {Lukin}},\ }\bibfield  {title} {\bibinfo {title}
  {High-sensitivity diamond magnetometer with nanoscale resolution},\ }\href
  {https://doi.org/10.1038/nphys1075} {\bibfield  {journal} {\bibinfo
  {journal} {Nature Physics}\ }\textbf {\bibinfo {volume} {4}},\ \bibinfo
  {pages} {810} (\bibinfo {year} {2008})}\BibitemShut {NoStop}%
\bibitem [{\citenamefont {Degen}\ \emph {et~al.}(2017)\citenamefont {Degen},
  \citenamefont {Reinhard},\ and\ \citenamefont
  {Cappellaro}}]{degen_quantum_2017}%
  \BibitemOpen
  \bibfield  {author} {\bibinfo {author} {\bibfnamefont {C.}~\bibnamefont
  {Degen}}, \bibinfo {author} {\bibfnamefont {F.}~\bibnamefont {Reinhard}},\
  and\ \bibinfo {author} {\bibfnamefont {P.}~\bibnamefont {Cappellaro}},\
  }\bibfield  {title} {\bibinfo {title} {Quantum sensing},\ }\href
  {https://doi.org/10.1103/RevModPhys.89.035002} {\bibfield  {journal}
  {\bibinfo  {journal} {Reviews of Modern Physics}\ }\textbf {\bibinfo {volume}
  {89}},\ \bibinfo {pages} {035002} (\bibinfo {year} {2017})}\BibitemShut
  {NoStop}%
\bibitem [{\citenamefont {Clarke}\ and\ \citenamefont
  {Wilhelm}(2008)}]{clarke_superconducting_2008}%
  \BibitemOpen
  \bibfield  {author} {\bibinfo {author} {\bibfnamefont {J.}~\bibnamefont
  {Clarke}}\ and\ \bibinfo {author} {\bibfnamefont {F.~K.}\ \bibnamefont
  {Wilhelm}},\ }\bibfield  {title} {\bibinfo {title} {Superconducting quantum
  bits},\ }\href {https://doi.org/10.1038/nature07128} {\bibfield  {journal}
  {\bibinfo  {journal} {Nature}\ }\textbf {\bibinfo {volume} {453}},\ \bibinfo
  {pages} {1031} (\bibinfo {year} {2008})}\BibitemShut {NoStop}%
\bibitem [{\citenamefont {Taylor}\ and\ \citenamefont
  {Bowen}(2016)}]{taylor_quantum_2016}%
  \BibitemOpen
  \bibfield  {author} {\bibinfo {author} {\bibfnamefont {M.~A.}\ \bibnamefont
  {Taylor}}\ and\ \bibinfo {author} {\bibfnamefont {W.~P.}\ \bibnamefont
  {Bowen}},\ }\bibfield  {title} {\bibinfo {title} {Quantum metrology and its
  application in biology},\ }\bibfield  {journal} {\bibinfo  {journal} {Physics
  Reports}\ }\bibinfo {series} {Quantum metrology and its application in
  biology},\ \textbf {\bibinfo {volume} {615}},\ \href
  {https://doi.org/10.1016/j.physrep.2015.12.002}
  {10.1016/j.physrep.2015.12.002} (\bibinfo {year} {2016})\BibitemShut
  {NoStop}%
\bibitem [{\citenamefont {Abouraddy}\ \emph {et~al.}(2002)\citenamefont
  {Abouraddy}, \citenamefont {Nasr}, \citenamefont {Saleh}, \citenamefont
  {Sergienko},\ and\ \citenamefont {Teich}}]{abouraddy_quantum-optical_2002}%
  \BibitemOpen
  \bibfield  {author} {\bibinfo {author} {\bibfnamefont {A.~F.}\ \bibnamefont
  {Abouraddy}}, \bibinfo {author} {\bibfnamefont {M.~B.}\ \bibnamefont {Nasr}},
  \bibinfo {author} {\bibfnamefont {B.~E.~A.}\ \bibnamefont {Saleh}}, \bibinfo
  {author} {\bibfnamefont {A.~V.}\ \bibnamefont {Sergienko}},\ and\ \bibinfo
  {author} {\bibfnamefont {M.~C.}\ \bibnamefont {Teich}},\ }\bibfield  {title}
  {\bibinfo {title} {Quantum-optical coherence tomography with dispersion
  cancellation},\ }\bibfield  {journal} {\bibinfo  {journal} {Physical Review
  A}\ }\textbf {\bibinfo {volume} {65}},\ \href
  {https://doi.org/10.1103/PhysRevA.65.053817} {10.1103/PhysRevA.65.053817}
  (\bibinfo {year} {2002})\BibitemShut {NoStop}%
\bibitem [{\citenamefont {Kolobov}(1999)}]{kolobov_spatial_1999}%
  \BibitemOpen
  \bibfield  {author} {\bibinfo {author} {\bibfnamefont {M.~I.}\ \bibnamefont
  {Kolobov}},\ }\bibfield  {title} {\bibinfo {title} {The spatial behavior of
  nonclassical light},\ }\href {https://doi.org/10.1103/RevModPhys.71.1539}
  {\bibfield  {journal} {\bibinfo  {journal} {Reviews of Modern Physics}\
  }\textbf {\bibinfo {volume} {71}},\ \bibinfo {pages} {1539} (\bibinfo {year}
  {1999})}\BibitemShut {NoStop}%
\bibitem [{\citenamefont {Clark}\ \emph {et~al.}(2021)\citenamefont {Clark},
  \citenamefont {Chekhova}, \citenamefont {Matthews}, \citenamefont {Rarity},\
  and\ \citenamefont {Oulton}}]{clark_special_2021}%
  \BibitemOpen
  \bibfield  {author} {\bibinfo {author} {\bibfnamefont {A.~S.}\ \bibnamefont
  {Clark}}, \bibinfo {author} {\bibfnamefont {M.}~\bibnamefont {Chekhova}},
  \bibinfo {author} {\bibfnamefont {J.~C.~F.}\ \bibnamefont {Matthews}},
  \bibinfo {author} {\bibfnamefont {J.~G.}\ \bibnamefont {Rarity}},\ and\
  \bibinfo {author} {\bibfnamefont {R.~F.}\ \bibnamefont {Oulton}},\ }\bibfield
   {title} {\bibinfo {title} {Special {Topic}: {Quantum} sensing with
  correlated light sources},\ }\href {https://doi.org/10.1063/5.0041043}
  {\bibfield  {journal} {\bibinfo  {journal} {Applied Physics Letters}\
  }\textbf {\bibinfo {volume} {118}},\ \bibinfo {pages} {060401} (\bibinfo
  {year} {2021})}\BibitemShut {NoStop}%
\bibitem [{\citenamefont {Rarity}\ \emph {et~al.}(1993)\citenamefont {Rarity},
  \citenamefont {Burnett}, \citenamefont {Tapster},\ and\ \citenamefont
  {Paschotta}}]{rarity_high_visibility_1993}%
  \BibitemOpen
  \bibfield  {author} {\bibinfo {author} {\bibfnamefont {J.~G.}\ \bibnamefont
  {Rarity}}, \bibinfo {author} {\bibfnamefont {J.}~\bibnamefont {Burnett}},
  \bibinfo {author} {\bibfnamefont {P.~R.}\ \bibnamefont {Tapster}},\ and\
  \bibinfo {author} {\bibfnamefont {R.}~\bibnamefont {Paschotta}},\ }\bibfield
  {title} {\bibinfo {title} {High visibility two photon interference in a
  single mode fibre interferometer},\ }\href
  {https://doi.org/10.1209/0295-5075/22/2/004} {\bibfield  {journal} {\bibinfo
  {journal} {Europhysics Letters (EPL)}\ }\textbf {\bibinfo {volume} {22}},\
  \bibinfo {pages} {95} (\bibinfo {year} {1993})}\BibitemShut {NoStop}%
\bibitem [{\citenamefont {Agrawal}(1995)}]{agrawal_nonlinear_1995}%
  \BibitemOpen
  \bibfield  {author} {\bibinfo {author} {\bibfnamefont {G.~P.}\ \bibnamefont
  {Agrawal}},\ }\href@noop {} {\emph {\bibinfo {title} {Nonlinear fiber
  optics}}},\ \bibinfo {edition} {2nd}\ ed.,\ Optics and photonics\ (\bibinfo
  {publisher} {Academic Press},\ \bibinfo {address} {San Diego},\ \bibinfo
  {year} {1995})\BibitemShut {NoStop}%
\bibitem [{\citenamefont {Fasel}\ \emph {et~al.}(2004)\citenamefont {Fasel},
  \citenamefont {Gisin}, \citenamefont {Ribordy},\ and\ \citenamefont
  {Zbinden}}]{fasel_quantum_2004}%
  \BibitemOpen
  \bibfield  {author} {\bibinfo {author} {\bibfnamefont {S.}~\bibnamefont
  {Fasel}}, \bibinfo {author} {\bibfnamefont {N.}~\bibnamefont {Gisin}},
  \bibinfo {author} {\bibfnamefont {G.}~\bibnamefont {Ribordy}},\ and\ \bibinfo
  {author} {\bibfnamefont {H.}~\bibnamefont {Zbinden}},\ }\bibfield  {title}
  {\bibinfo {title} {Quantum key distribution over 30 km of standard fiber
  using energy-time entangled photon pairs: a comparison of two chromatic
  dispersion reduction methods},\ }\href
  {https://doi.org/10.1140/epjd/e2004-00080-8} {\bibfield  {journal} {\bibinfo
  {journal} {The European Physical Journal D}\ }\textbf {\bibinfo {volume}
  {30}},\ \bibinfo {pages} {143} (\bibinfo {year} {2004})}\BibitemShut
  {NoStop}%
\bibitem [{The(2009)}]{Thevenaz}%
  \BibitemOpen
  \href@noop {} {\emph {\bibinfo {title} {Proc.23rd Int. Symposium on
  Distributed Computing}}},\ \bibinfo {series} {Lecture Notes in Computer
  Science}, Vol.\ \bibinfo {volume} {5805}\ (\bibinfo  {publisher} {Springer},\
  \bibinfo {address} {Berlin, Germany},\ \bibinfo {year} {2009})\BibitemShut
  {NoStop}%
\bibitem [{\citenamefont {Labonte}\ \emph {et~al.}(2006)\citenamefont
  {Labonte}, \citenamefont {Roy}, \citenamefont {Pagnoux}, \citenamefont
  {Louradour}, \citenamefont {Restoin}, \citenamefont {Mélin},\ and\
  \citenamefont {Burov}}]{Disp}%
  \BibitemOpen
  \bibfield  {author} {\bibinfo {author} {\bibfnamefont {L.}~\bibnamefont
  {Labonte}}, \bibinfo {author} {\bibfnamefont {P.}~\bibnamefont {Roy}},
  \bibinfo {author} {\bibfnamefont {D.}~\bibnamefont {Pagnoux}}, \bibinfo
  {author} {\bibfnamefont {F.}~\bibnamefont {Louradour}}, \bibinfo {author}
  {\bibfnamefont {C.}~\bibnamefont {Restoin}}, \bibinfo {author} {\bibfnamefont
  {G.}~\bibnamefont {Mélin}},\ and\ \bibinfo {author} {\bibfnamefont
  {E.}~\bibnamefont {Burov}},\ }\bibfield  {title} {\bibinfo {title}
  {Experimental and numerical analysis of the chromatic dispersion dependence
  upon the actual profile of small core microstructured fibres},\ }\href
  {https://doi.org/10.1088/1464-4258/8/11/001} {\bibfield  {journal} {\bibinfo
  {journal} {Journal of Optics A: Pure and Applied Optics}\ }\textbf {\bibinfo
  {volume} {8}},\ \bibinfo {pages} {933} (\bibinfo {year} {2006})}\BibitemShut
  {NoStop}%
\bibitem [{\citenamefont {Kaiser}\ \emph {et~al.}(2018)\citenamefont {Kaiser},
  \citenamefont {Vergyris}, \citenamefont {Aktas}, \citenamefont {Babin},
  \citenamefont {Labonte},\ and\ \citenamefont
  {Tanzilli}}]{kaiser_quantum_2018}%
  \BibitemOpen
  \bibfield  {author} {\bibinfo {author} {\bibfnamefont {F.}~\bibnamefont
  {Kaiser}}, \bibinfo {author} {\bibfnamefont {P.}~\bibnamefont {Vergyris}},
  \bibinfo {author} {\bibfnamefont {D.}~\bibnamefont {Aktas}}, \bibinfo
  {author} {\bibfnamefont {C.}~\bibnamefont {Babin}}, \bibinfo {author}
  {\bibfnamefont {L.}~\bibnamefont {Labonte}},\ and\ \bibinfo {author}
  {\bibfnamefont {S.}~\bibnamefont {Tanzilli}},\ }\bibfield  {title} {\bibinfo
  {title} {Quantum enhancement of accuracy and precision in optical
  interferometry},\ }\href {https://doi.org/10.1038/lsa.2017.163} {\bibfield
  {journal} {\bibinfo  {journal} {Light: Science \& Applications}\ }\textbf
  {\bibinfo {volume} {7}},\ \bibinfo {pages} {17163} (\bibinfo {year}
  {2018})}\BibitemShut {NoStop}%
\bibitem [{\citenamefont {Franson}(1989)}]{franson_bell_1989}%
  \BibitemOpen
  \bibfield  {author} {\bibinfo {author} {\bibfnamefont {J.~D.}\ \bibnamefont
  {Franson}},\ }\bibfield  {title} {\bibinfo {title} {Bell inequality for
  position and time},\ }\href {https://doi.org/10.1103/PhysRevLett.62.2205}
  {\bibfield  {journal} {\bibinfo  {journal} {Physical Review Letters}\
  }\textbf {\bibinfo {volume} {62}},\ \bibinfo {pages} {2205} (\bibinfo {year}
  {1989})}\BibitemShut {NoStop}%
\bibitem [{\citenamefont {Aktas}\ \emph {et~al.}(2016)\citenamefont {Aktas},
  \citenamefont {Fedrici}, \citenamefont {Kaiser}, \citenamefont {Lunghi},
  \citenamefont {Labonte},\ and\ \citenamefont
  {Tanzilli}}]{aktas_entanglement_2016}%
  \BibitemOpen
  \bibfield  {author} {\bibinfo {author} {\bibfnamefont {D.}~\bibnamefont
  {Aktas}}, \bibinfo {author} {\bibfnamefont {B.}~\bibnamefont {Fedrici}},
  \bibinfo {author} {\bibfnamefont {F.}~\bibnamefont {Kaiser}}, \bibinfo
  {author} {\bibfnamefont {T.}~\bibnamefont {Lunghi}}, \bibinfo {author}
  {\bibfnamefont {L.}~\bibnamefont {Labonte}},\ and\ \bibinfo {author}
  {\bibfnamefont {S.}~\bibnamefont {Tanzilli}},\ }\bibfield  {title} {\bibinfo
  {title} {Entanglement distribution over 150 km in wavelength division
  multiplexed channels for quantum cryptography},\ }\href
  {https://doi.org/https://doi.org/10.1002/lpor.201500258} {\bibfield
  {journal} {\bibinfo  {journal} {Laser \& Photonics Reviews}\ }\textbf
  {\bibinfo {volume} {10}},\ \bibinfo {pages} {451} (\bibinfo {year}
  {2016})}\BibitemShut {NoStop}%
\bibitem [{\citenamefont {Autebert}\ \emph {et~al.}(2015)\citenamefont
  {Autebert}, \citenamefont {Bruno}, \citenamefont {Martin}, \citenamefont
  {Lemaître}, \citenamefont {Gomez~Carbonell}, \citenamefont {Favero},
  \citenamefont {Leo}, \citenamefont {Zbinden},\ and\ \citenamefont
  {Ducci}}]{Autebert}%
  \BibitemOpen
  \bibfield  {author} {\bibinfo {author} {\bibfnamefont {C.}~\bibnamefont
  {Autebert}}, \bibinfo {author} {\bibfnamefont {N.}~\bibnamefont {Bruno}},
  \bibinfo {author} {\bibfnamefont {A.}~\bibnamefont {Martin}}, \bibinfo
  {author} {\bibfnamefont {A.}~\bibnamefont {Lemaître}}, \bibinfo {author}
  {\bibfnamefont {C.}~\bibnamefont {Gomez~Carbonell}}, \bibinfo {author}
  {\bibfnamefont {I.}~\bibnamefont {Favero}}, \bibinfo {author} {\bibfnamefont
  {G.}~\bibnamefont {Leo}}, \bibinfo {author} {\bibfnamefont {H.}~\bibnamefont
  {Zbinden}},\ and\ \bibinfo {author} {\bibfnamefont {S.}~\bibnamefont
  {Ducci}},\ }\bibfield  {title} {\bibinfo {title} {Integrated algaas source of
  highly indistinguishable and energy-time entangled photons},\ }\href
  {https://doi.org/10.1364/OPTICA.3.000143} {\bibfield  {journal} {\bibinfo
  {journal} {Optica}\ }\textbf {\bibinfo {volume} {3}} (\bibinfo {year}
  {2015})}\BibitemShut {NoStop}%
\bibitem [{\citenamefont {Oser}\ \emph {et~al.}(2020)\citenamefont {Oser},
  \citenamefont {Tanzilli}, \citenamefont {Mazeas}, \citenamefont
  {Alonso~Ramos}, \citenamefont {Le~Roux}, \citenamefont {Sauder},
  \citenamefont {Hua}, \citenamefont {Alibart}, \citenamefont {Vivien},
  \citenamefont {Cassan},\ and\ \citenamefont
  {Labonteeeee}}]{oser_high_quality_2020}%
  \BibitemOpen
  \bibfield  {author} {\bibinfo {author} {\bibfnamefont {D.}~\bibnamefont
  {Oser}}, \bibinfo {author} {\bibfnamefont {S.}~\bibnamefont {Tanzilli}},
  \bibinfo {author} {\bibfnamefont {F.}~\bibnamefont {Mazeas}}, \bibinfo
  {author} {\bibfnamefont {C.}~\bibnamefont {Alonso~Ramos}}, \bibinfo {author}
  {\bibfnamefont {X.}~\bibnamefont {Le~Roux}}, \bibinfo {author} {\bibfnamefont
  {G.}~\bibnamefont {Sauder}}, \bibinfo {author} {\bibfnamefont
  {X.}~\bibnamefont {Hua}}, \bibinfo {author} {\bibfnamefont {O.}~\bibnamefont
  {Alibart}}, \bibinfo {author} {\bibfnamefont {L.}~\bibnamefont {Vivien}},
  \bibinfo {author} {\bibfnamefont {E.}~\bibnamefont {Cassan}},\ and\ \bibinfo
  {author} {\bibfnamefont {L.}~\bibnamefont {Labonteeeee}},\ }\bibfield
  {title} {\bibinfo {title} {High quality photonic entanglement out of a
  stand-alone silicon chip},\ }\href
  {https://doi.org/10.1038/s41534-020-0263-7} {\bibfield  {journal} {\bibinfo
  {journal} {npj Quantum Information}\ }\textbf {\bibinfo {volume} {6}},\
  \bibinfo {pages} {31} (\bibinfo {year} {2020})}\BibitemShut {NoStop}%
\bibitem [{\citenamefont {Riazi}\ \emph {et~al.}(2019)\citenamefont {Riazi},
  \citenamefont {Zhu}, \citenamefont {Chen}, \citenamefont {Gladyshev},
  \citenamefont {Kazansky},\ and\ \citenamefont {Qian}}]{Riazi:19}%
  \BibitemOpen
  \bibfield  {author} {\bibinfo {author} {\bibfnamefont {A.}~\bibnamefont
  {Riazi}}, \bibinfo {author} {\bibfnamefont {E.~Y.}\ \bibnamefont {Zhu}},
  \bibinfo {author} {\bibfnamefont {C.}~\bibnamefont {Chen}}, \bibinfo {author}
  {\bibfnamefont {A.~V.}\ \bibnamefont {Gladyshev}}, \bibinfo {author}
  {\bibfnamefont {P.~G.}\ \bibnamefont {Kazansky}},\ and\ \bibinfo {author}
  {\bibfnamefont {L.}~\bibnamefont {Qian}},\ }\bibfield  {title} {\bibinfo
  {title} {Alignment-free dispersion measurement with interfering biphotons},\
  }\href {https://doi.org/10.1364/OL.44.001484} {\bibfield  {journal} {\bibinfo
   {journal} {Opt. Lett.}\ }\textbf {\bibinfo {volume} {44}},\ \bibinfo {pages}
  {1484} (\bibinfo {year} {2019})}\BibitemShut {NoStop}%
\bibitem [{\citenamefont {Scarani}\ \emph {et~al.}(2009)\citenamefont
  {Scarani}, \citenamefont {Bechmann-Pasquinucci}, \citenamefont {Cerf},
  \citenamefont {Dušek}, \citenamefont {Lütkenhaus},\ and\ \citenamefont
  {Peev}}]{scarani_security_2009}%
  \BibitemOpen
  \bibfield  {author} {\bibinfo {author} {\bibfnamefont {V.}~\bibnamefont
  {Scarani}}, \bibinfo {author} {\bibfnamefont {H.}~\bibnamefont
  {Bechmann-Pasquinucci}}, \bibinfo {author} {\bibfnamefont {N.~J.}\
  \bibnamefont {Cerf}}, \bibinfo {author} {\bibfnamefont {M.}~\bibnamefont
  {Dušek}}, \bibinfo {author} {\bibfnamefont {N.}~\bibnamefont
  {Lütkenhaus}},\ and\ \bibinfo {author} {\bibfnamefont {M.}~\bibnamefont
  {Peev}},\ }\bibfield  {title} {\bibinfo {title} {The security of practical
  quantum key distribution},\ }\href
  {https://doi.org/10.1103/RevModPhys.81.1301} {\bibfield  {journal} {\bibinfo
  {journal} {Reviews of Modern Physics}\ }\textbf {\bibinfo {volume} {81}},\
  \bibinfo {pages} {1301} (\bibinfo {year} {2009})}\BibitemShut {NoStop}%
\bibitem [{\citenamefont {Zhong}\ and\ \citenamefont
  {Wong}(2013)}]{zhong_nonlocal_2013}%
  \BibitemOpen
  \bibfield  {author} {\bibinfo {author} {\bibfnamefont {T.}~\bibnamefont
  {Zhong}}\ and\ \bibinfo {author} {\bibfnamefont {F.~N.~C.}\ \bibnamefont
  {Wong}},\ }\bibfield  {title} {\bibinfo {title} {Nonlocal cancellation of
  dispersion in {Franson} interferometry},\ }\bibfield  {journal} {\bibinfo
  {journal} {Physical Review A}\ }\textbf {\bibinfo {volume} {88}},\ \href
  {https://doi.org/10.1103/PhysRevA.88.020103} {10.1103/PhysRevA.88.020103}
  (\bibinfo {year} {2013})\BibitemShut {NoStop}%
\bibitem [{\citenamefont {Zhang}\ \emph {et~al.}(2014)\citenamefont {Zhang},
  \citenamefont {Mower}, \citenamefont {Englund}, \citenamefont {Wong},\ and\
  \citenamefont {Shapiro}}]{zhang_unconditional_2014}%
  \BibitemOpen
  \bibfield  {author} {\bibinfo {author} {\bibfnamefont {Z.}~\bibnamefont
  {Zhang}}, \bibinfo {author} {\bibfnamefont {J.}~\bibnamefont {Mower}},
  \bibinfo {author} {\bibfnamefont {D.}~\bibnamefont {Englund}}, \bibinfo
  {author} {\bibfnamefont {F.~N.}\ \bibnamefont {Wong}},\ and\ \bibinfo
  {author} {\bibfnamefont {J.~H.}\ \bibnamefont {Shapiro}},\ }\bibfield
  {title} {\bibinfo {title} {Unconditional {Security} of {Time}-{Energy}
  {Entanglement} {Quantum} {Key} {Distribution} {Using} {Dual}-{Basis}
  {Interferometry}},\ }\href {https://doi.org/10.1103/PhysRevLett.112.120506}
  {\bibfield  {journal} {\bibinfo  {journal} {Physical Review Letters}\
  }\textbf {\bibinfo {volume} {112}},\ \bibinfo {pages} {120506} (\bibinfo
  {year} {2014})}\BibitemShut {NoStop}%
\bibitem [{\citenamefont {Leviton}\ and\ \citenamefont
  {Frey}(2008)}]{leviton_temperature-dependent_2008}%
  \BibitemOpen
  \bibfield  {author} {\bibinfo {author} {\bibfnamefont {D.~B.}\ \bibnamefont
  {Leviton}}\ and\ \bibinfo {author} {\bibfnamefont {B.~J.}\ \bibnamefont
  {Frey}},\ }\href {https://doi.org/10.48550/arXiv.0805.0091} {\bibinfo {title}
  {Temperature-dependent absolute refractive index measurements of synthetic
  fused silica}} (\bibinfo {year} {2008})\BibitemShut {NoStop}%
\bibitem [{\citenamefont {Corning}(2000)}]{Corning}%
  \BibitemOpen
  \bibfield  {author} {\bibinfo {author} {\bibnamefont {Corning}},\ }\href@noop
  {} {\bibinfo {title} {Corning smf-28 ull optical fiber}} (\bibinfo {year}
  {2000})\BibitemShut {NoStop}%
\bibitem [{\citenamefont {Baker}\ \emph {et~al.}(2014)\citenamefont {Baker},
  \citenamefont {Lu},\ and\ \citenamefont
  {Bao}}]{baker_chromatic-dispersion_2014}%
  \BibitemOpen
  \bibfield  {author} {\bibinfo {author} {\bibfnamefont {C.}~\bibnamefont
  {Baker}}, \bibinfo {author} {\bibfnamefont {Y.}~\bibnamefont {Lu}},\ and\
  \bibinfo {author} {\bibfnamefont {X.}~\bibnamefont {Bao}},\ }\bibfield
  {title} {\bibinfo {title} {Chromatic-dispersion measurement by modulation
  phase-shift method using a {Kerr} phase-interrogator},\ }\href
  {https://doi.org/10.1364/OE.22.022314} {\bibfield  {journal} {\bibinfo
  {journal} {Optics Express}\ }\textbf {\bibinfo {volume} {22}},\ \bibinfo
  {pages} {22314} (\bibinfo {year} {2014})}\BibitemShut {NoStop}%
\bibitem [{\citenamefont {Sucbei}\ and\ \citenamefont
  {Kim}(2009)}]{sucbei_moon_reflectometric_2009}%
  \BibitemOpen
  \bibfield  {author} {\bibinfo {author} {\bibfnamefont {M.}~\bibnamefont
  {Sucbei}}\ and\ \bibinfo {author} {\bibfnamefont {D.}~\bibnamefont {Kim}},\
  }\bibfield  {title} {\bibinfo {title} {Reflectometric fiber dispersion
  measurement using a supercontinuum pulse source},\ }\href
  {https://doi.org/10.1109/LPT.2009.2025381} {\bibfield  {journal} {\bibinfo
  {journal} {IEEE Photonics Technology Letters}\ }\textbf {\bibinfo {volume}
  {21}},\ \bibinfo {pages} {1262} (\bibinfo {year} {2009})}\BibitemShut
  {NoStop}%
\bibitem [{\citenamefont {Grosz}\ \emph {et~al.}(2014)\citenamefont {Grosz},
  \citenamefont {Kovacs}, \citenamefont {Kiss},\ and\ \citenamefont
  {Szipocs}}]{grosz_measurement_2014}%
  \BibitemOpen
  \bibfield  {author} {\bibinfo {author} {\bibfnamefont {T.}~\bibnamefont
  {Grosz}}, \bibinfo {author} {\bibfnamefont {A.~P.}\ \bibnamefont {Kovacs}},
  \bibinfo {author} {\bibfnamefont {M.}~\bibnamefont {Kiss}},\ and\ \bibinfo
  {author} {\bibfnamefont {R.}~\bibnamefont {Szipocs}},\ }\bibfield  {title}
  {\bibinfo {title} {Measurement of higher order chromatic dispersion in a
  photonic bandgap fiber},\ }\href {https://doi.org/10.1364/AO.53.001929}
  {\bibfield  {journal} {\bibinfo  {journal} {Applied Optics}\ }\textbf
  {\bibinfo {volume} {53}},\ \bibinfo {pages} {1929} (\bibinfo {year}
  {2014})}\BibitemShut {NoStop}%
\end{thebibliography}%

\end{document}